# Image-derived generative modeling of pseudo-macromolecular structures – towards the statistical assessment of Electron CryoTomography template matching

Kai Wen Wang[1], Xiangrui Zeng[2], Xiaodan Liang[3], Zhiguang Huo[4], Eric P. Xing[3], and Min Xu[2,*]

[1]*Computer Science Department, Carnegie Mellon University, Pittsburgh, 15213, USA.*
[2]*Computational Biology Department, Carnegie Mellon University, Pittsburgh, 15213, USA.*
[3]*Machine Learning Department, Carnegie Mellon University, Pittsburgh, 15213, USA.*
[4]*Department of Biostatistics, University of Florida, Gainesville, 32611, USA.*
[*]*Corresponding author email: mxu1@cs.cmu.edu*

## Abstract

Cellular Electron CryoTomography (CECT) is a 3D imaging technique that captures information about the structure and spatial organization of macromolecular complexes within single cells, in near-native state and at sub-molecular resolution. Although template matching is often used to locate macromolecules in a CECT image, it is insufficient as it only measures the relative structural similarity. Therefore, it is preferable to assess the statistical credibility of the decision through hypothesis testing, requiring many templates derived from a diverse population of macromolecular structures. Due to the very limited number of known structures, we need a generative model to efficiently and reliably sample pseudo-structures from the complex distribution of macromolecular structures. To address this challenge, we propose a novel image-derived approach for performing hypothesis testing for template matching by constructing generative models using the generative adversarial network. Finally, we conducted hypothesis testing experiments for template matching on both simulated and experimental subtomograms, allowing us to conclude the identity of subtomograms with high statistical credibility and significantly reducing false positives.

# 1 Introduction

The cell is the basic structural and functional unit of all living organisms. Biochemical processes of the living cell are often catalyzed by tiny cellular machines called macromolecular complexes. To fully understand these cellular processes, it is extremely helpful to systematically extract the structure and spatial organization of macromolecular complexes in single cells. Cellular Electron CryoTomography (CECT) [2] is a powerful 3D imaging tool that enables the study of sub-cellular structures at near-native state and in sub-molecular resolution. However, the quality of the reconstructed CECT images (a.k.a. tomograms) suffer from many current imaging limitations, such as low signal-to-noise ratio (SNR), missing wedge and limited number of angular samples, to a point where interpretation by visual inspection is impractical. As such, locating instances of macromolecular complexes inside tomograms has remained an extremely challenging computer vision problem [7]. A popular method for this task has been template matching. However, this approach is inadequate as it lacks an absolute metric. We propose a novel image-derived, Monte Carlo approach for performing hypothesis testing for template matching using generative models.

Let $T(C)$ denote the template of complex $C$ and let $\mathscr{S}_{known}$ denote the set of known macromolecular complexes. Given a subtomogram $P$, a cubic sub-volume of a tomogram containing a single macromolecule, and a template $T(C), C \in \mathscr{S}_{known}$, we can calculate the cross-correlation $c(P, T(C))$, a relative measure of the structural similarity of $P$ to $T(C)$, as follows [10]:

1. Rigid alignment of $T(C)$ against $P$ using a fast alignment method [34, 10].

2. Calculate $c(T(C), P)$ using Pearson correlation with missing wedge compensation [8].

Template matching decides whether $P$ contains an identical macromolecular structure as $C$ (we denote a match as $P = T(C)$). Existing template matching procedures either use a single chosen threshold for $c$ or take the template most similar to $P$ in terms of the highest cross-correlation. However, as $c$ is only a relative measure, these procedures are neither rigorous nor statistically meaningful. Instead, hypothesis testing is preferred for a quantitative assessment of the statistical credibility.

In this paper, we propose a statistically rigorous treatment of template matching, performed in two steps. Given a subtomogram $P$ and a set of known templates $\mathscr{S}_{known}$, we follow the two-step process:

(Step 1) Determine the complex of interest $C_I = \arg\max_{C \in \mathscr{S}_{known}} c(P, T(C))$.

(Step 2) Perform hypothesis test for $P = T(C_I)$ and calculate the p-value.

We select the complex of interest $C_I$ as the complex with the highest alignment score. To formally establish the hypothesis testing procedure for $P = T(C_I)$, we specify the null hypothesis $H_0$ and the alternative hypothesis $H_A$ as the following:

$$H_0 : P \neq T(C_I) \qquad\qquad H_A : P = T(C_I) \qquad\qquad (1)$$

Let $C_0$ be a random macromolecules drawn from the structural distribution of macromolecules. We utilize $c(P, T(C_0))$ as test statistics to evaluate the hypothesis testing procedure. The statistical credibility is assessed by the p-value, which is the probability of obtaining $c(P, T(C_0))$ at least as extreme as $c(P, T(C_I))$, given that $H_0 : P \neq T(C_I)$ is true. The lower the p-value, the stronger the evidence is against $H_0$, which gives more statistical credibility that $H_A : P = T(C_I)$ is true.

The challenge of performing an accurate hypothesis test is that only a small number of macromolecules have known structures. In addition, the structural distribution of macromolecules is highly complex and cannot be easily approximated. Across space and time, macromolecules typically adopt different conformations as part of their function and dynamically interact with other macromolecules [36]. Across different species and cell types, the majority of macromolecules are still unknown [14]. Since there is no parametric distribution form of the test statistic $c(P, T(C_0))$, we propose to use Monte Carlo simulation approach to calculate empirical p-values under $H_0$ [16].

We first construct a generative model to learn the structural distribution $f_0$ from a collection of known complexes $\mathscr{S}_{known}$ (such as from Protein DataBank). To calculate an empirical p-value, we randomly sample pseudo-macromolecular complexes $C_0 \sim f_0$ to derive a Monte Carlo empirical distribution of the test statistics $c(P, T(C_0))$ under the null. In this paper, let $\mathscr{S}_{pseudo}$ denote the set of pseudo-complexes. We emphasize that $\mathscr{S}_{pseudo}$ and $\mathscr{S}_{known}$ have important differences:

- The identity of $C \in \mathscr{S}_{known}$ is known and $T(C)$ is used to identify the macromolecule in a subtomogram through the method of template matching.

- The identity of $\tilde{C} \in \mathscr{S}_{pseudo}$ is unknown and $T(\tilde{C})$ is only used for hypothesis testing.

To learn $f_0$ , we use the generative adversarial network (GAN) to produce pseudo-macromolecular complexes represented as 3D gray-scale images known as density maps. As a recent advancement in unsupervised deep learning, the GAN learns the distribution of training images on the image manifold and generates highly realistic images [12]. The training process of the GAN is akin to a minimax game between two neural network adversaries, the generator and the discriminator. The generator seeks to improve its output images by minimizing the discriminator's classification accuracy while the discriminator seeks to maximize its accuracy. After training, the generator can produce diverse and realistic images from the distribution of the original training images. To our knowledge, no method exists for constructing such generative models for template search for CECT data.

**Our contributions are summarized as follows:**

(i) A novel approach for statistical assessing template matching through hypothesis testing to calculate Monte Carlo empirical p-values.

(ii) An approach for generating the density maps of pseudo-macromolecular complexes, using the 3D Deep Convolutional Wasserstein GAN (3D-WGAN). We showed that our model is able to capture the shape manifold of macromolecules, and sample realistic and diverse pseudo-complexes from the structural distribution of macromolecules.

## 2 Methods

### 2.1 3D Deep Convolutional Wasserstein GAN (3D-WGAN)

Our generative approach combines the 3D-GAN [33] with the Improved Wasserstein GAN [13], so we name our model 3D Deep Convolutional Wasserstein GAN (3D-WGAN) [1]. The network architecture of our 3D-WGAN is presented in Figure 1. Inspired by [33], our network generator and the discriminator are implemented each with four convolution layers of stride 2 and kernel size $4^3$, which we chose to be a factor of 64 to reduce the checkerboard artifact [26]. We found that using half as many filters as [33] was sufficient for the GAN to stabilize at producing good results. The input to the generator is a Gaussian random vector in $\mathbb{R}^{100}$. In the hidden layers, we used the LeakyReLU activation with $\alpha = 0.2$. Following [1, 13], we only used Batch Normalization [17] in the generator.

A common problem for GANs with especially limited data is mode collapse, which occurs when the generator only produces structures with very low diversity. We adopted the minibatch discrimination layer in the discriminator (see [28]) to reduce the collapse of the generator by penalizing low-entropy generators. This layer allows the discriminator to observe many samples at once, so that it can also take entropy of a batch of samples into account when deciding between real and pseudo images.

---

[1] Background details about GAN, as well as specifics on the convolutional layer, batch normalization layer, activation layers and the minibatch discrimination layer can be found in Supplementary Section 1.4

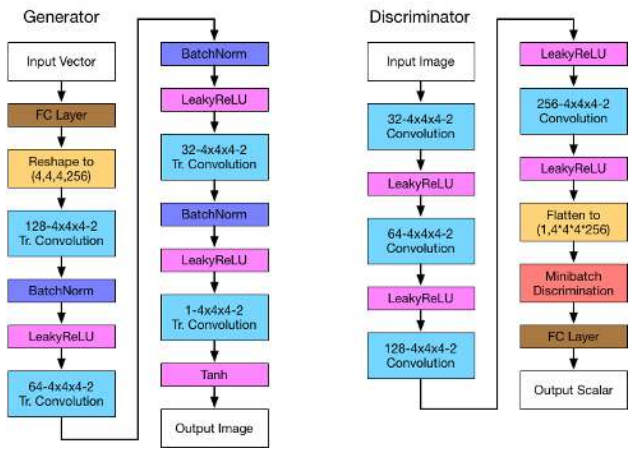

Figure 1: The network architecture for both the generator and discriminator of the 3D-WGAN. Each convolution layer is labeled in the format $N - K \times K \times K - S$, signifying $N$ filters with kernel size $K^3$ and stride $S$.

## 2.2 Training the 3D-WGAN

In our experiments, each macromolecular complex was represented as a density map (i.e. 3D gray-scale image) with $64^3$ voxels and 0.6nm pixel size. We constructed our dataset with 15 experimental macromolecular complexes that are diverse in shape and size (see Figure 3B). To prevent overfitting to a specific orientation or structure, we performed data augmentation and rotated each structure 600 times for a total of 9000 training structures. As the 3D convolution operation is not rotation invariant, this data augmentation improves the training of the 3D-WGAN. We trained the 3D-WGAN using a batch size of 64 and the Adam optimizer with $\beta_1 = 0.5$, $\beta_2 = 0.99$, and the learning rate as 0.0001, shown to be successful in previous works [27, 33]. Following [13], we trained the discriminator ten times as often as the generator, and we used a gradient penalty of 10.

## 2.3 Sampling pseudo-complexes far away from $C_I$

Under the null hypothesis that $P \neq T(C_I)$, the instance of $C_I$ should be not be covered by the distribution of pseudo-complexes. This procedure also reduces the chance of sampling pseudo-complexes so similar to $C_I$ that they could be viewed as copies of $C_I$ in the hypothesis test. With the following three-step procedure, we could sample pseudo-complexes $C_0$ from $f_0$ to be "far away" from $C_I$ in the latent representation of $f_0$:

1. Regressor (a.k.a. Inverse Generator): We trained a regressor for the inverse map of the generator, using a 3D extension of the AutoEncoder GAN model (see [21]). Instead of using cross-entropy loss, we used the sum squared error of the reconstructed images since our images were not normalized to $(0, 1)$ by a sigmoid. The network architecture of the regressor is the exact mirror image of the 3D-WGAN generator.

2. Kernel Density Estimator (KDE): We trained a KDE to learn the distribution $\mathscr{E}$ of the latent representation of $C_I$, given by the output of the regressor for 300 random rotations of $C_I$. The KDE's bandwidth was determined using 3-fold cross validation.

3. Bayes Classifier: Let $\pi$ ($\pi \ll 1$) be the prior probability of pseudo-complexes, which can be estimated from the data. Denote $\mathscr{N}$ as the standard 100-dimensional multivariate Gaussian. The decision boundary to distinguish the distribution of pseudo-complexes and $C_I$ is given by $Pr(C \in S_{pseudo}|P) =$

$Pr(C \in S_{known}|P)$. According to Bayes rule, the rejection region can be written as:

$$\mathscr{R} = \{G(P) : P \in \mathbb{R}^{100}, N(P) < \pi \cdot \mathscr{E}(P)\}$$

where $G(P)$ denotes generated pseudo-complex from $P$. When we sample points from $N$ for input to the 3D-WGAN, we reject the complexes which are members of $\mathscr{R}$.

## 2.4 Monte Carlo approach for evaluating the statistical credibility of template matching

After determining the complex of interest $C_I$ for Step 1, we perform a statistical assessment of $P = T(C_I)$ for Step 2 by calculating an empirical p-value using pseudo-macromolecular complexes generated by 3D-WGAN. The true p-value $p$ is the probability of obtaining results at least as extreme as the observed $c(P, T(C_I))$ given the null hypothesis $H_0 : P \neq T(C_I)$ is true. Since the distribution of $C_0$ under $f_0$ is unknown, we use a Monte Carlo simulation to obtain an unbiased empirical p-value $\hat{p}$ by ranking the observed test statistic $c(P, T(C_I))$ amongst the alignment scores of complexes sampled from the learned distribution of the 3D-WGAN. By the strong law of large numbers, our empirical estimate of the p-value converges almost surely to the true p-value as the number of Monte Carlo samples $B \to \infty$.

$$p = \mathbb{E}_{H_0 : P \neq T(C_I), C_0 \sim f_0} [\mathbb{I}\{c(P, T(C_I)) \leq c(P, T(C_0))\}]|C_0 \notin \mathscr{R} \tag{2}$$

$$= \Pr(c(P, T(C_I)) \leq c(P, T(C_0))|C_0 \notin \mathscr{R}) \tag{3}$$

$$\hat{p} = B^{-1} \sum_{b=1}^{B} \mathbb{I}[c(P, T(C_I)) \leq c(P, T(C_0^{(b)}))] \xrightarrow[B \to \infty]{a.s.} p \tag{4}$$

where $\{C_0^{(b)}\}_{1 \leq b \leq B}$ are Monte Carlo samples from $f_0$ and out of the rejection region $\mathscr{R}$, specified in Section 2.3.

The statistical credibility of the decision is measured by the p-value: the smaller the p-value, the more statistical credibility we have to reject the null and to support the alternative. Therefore, when $c(P, T(C_I))$ is ranked in the highest 1% of the templates used for template matching (i.e. $\hat{p} \leq 0.01$), we could conclude $P = T(C_I)$ with high confidence. We note here that in CECT template matching tasks, there are usually a large number of hypothesis tests for different subtomograms matching, which will consequently generate a large number of p-values. Multiple comparison is also suggested to be adjusted to control false discovery rate [5] to provide a stringent statistical criteria.

# 3 Results

## 3.1 Examples of pseudo-macromolecular complexes

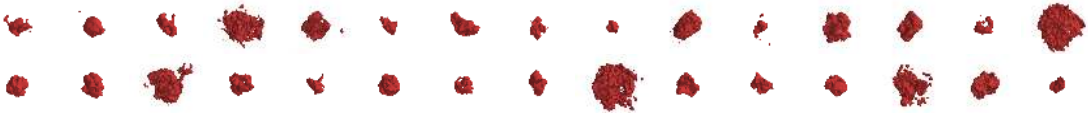

Figure 2: Random pseudo-macromolecular complexes generated with the 3D-WGAN.

Randomly selected pseudo-complexes from the 3D-WGAN are shown in Figure 2. Figure 3A shows the nearest neighbors of pseudo-complexes from the set of known complexes $\mathscr{S}_{known}$. Most pseudo-complexes exhibit similar structure as their nearest neighbor in $\mathscr{S}_{known}$ and resemble the same macromolecular complex. Figure 3B shows the nearest neighbors of known complexes from a set of 10,000 generated pseudo-complexes. We defined the metric between complexes as the $L_2$ norm on the fully-connected layer of the discriminator, which is a high-level feature representation of each complex [33]. Even as nearest neighbors of known complexes, the pseudo-complexes in Figure 3 are visually not identical to the known complexes. Non-rigid differences between the pseudo-complexes and the known complexes make these pseudo-complexes good candidates for hypothesis testing. This shows that our model can produce meaningful pseudo-complexes that have recognizable structural similarities to the training structures. A practical advantage to using the 3D-WGAN is that the generation of pseudo-complexes can be significantly sped up with a GPU. For example, our GTX 1080 Ti GPU could generate 10,000 structures in about 20s (0.002s/image).

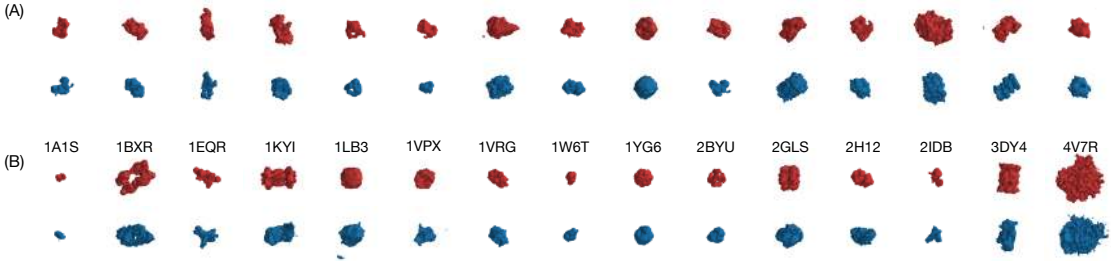

Figure 3: Let $C_{ij}$ denote the structure in the ith row and the jth column. (A) Each $C_{1j} \in \mathscr{S}_{pseudo}$, and $C_{2j} \in \mathscr{S}_{known}$ is the most similar structure to $C_{1j}$ in $\mathscr{S}_{known}$. (B) Similarly, Each $C_{3j} \in \mathscr{S}_{known}$, listed according to their PDB ID in Supplementary Section 1.3, and $C_{4j} \in \mathscr{S}_{pseudo}$ is the most similar structure to $C_{3j}$ in $\mathscr{S}_{pseudo}$. Blue color is represents the nearest shapes.

## 3.2 Statistical assessment of template matching on simulated subtomograms

Template matching is applied to real data to decide whether an unidentified subtomogram $P$ contains a macromolecule of known structure $C$ from the set of known structures $\mathscr{S}_{known}$. To mimic this setting, we simulate subtomograms using known complexes. For any fixed $C \in \mathscr{S}_{known}$, we simulate a subtomogram $P_C$ containing $C$. The hypothesis test consists of 985 randomly generated pseudo-templates $T(\tilde{C}) \in \mathscr{S}_{pseudo}$ and 15 templates of known structure $T(C) \in \mathscr{S}_{known}{}^2$. We performed template matching on $P$ following the two-step process in Section 1. A simulated test is successful if both conditions are satisfied:

(Cond. 1) <u>Highest alignment score:</u> $c(P_C, T(C)) \geq c(P_C, T(C'))$, $\forall C' \in \mathscr{S}_{known}$. (i.e. $C_I = C$)

(Cond. 2) <u>Low p-value:</u> The hypothesis test for $P = T(C_I)$ has p-value $\hat{p} \leq 0.01$.

We performed a simulated hypothesis test for each complex of known structure for a total of 15 tests and achieved an average success rate of $12/15 = 80\%$, indicating high power of our hypothesis testing procedure. Two successful cases are shown in Figure 4. Two failure cases are shown in Supplementary Section 1.5 and the rest of the results are shown in Supplementary Section 1.6.

From the smooth distribution of histograms in Figure 4, we can deduce that our generative model samples diverse and realistic pseudo-complexes from $f_0$, instead of reproducing the known complexes. Since if the pseudo-complexes were largely replicated known complexes, the plots would result in the histograms clustering only around the vertical lines.

## 3.3 Statistical assessment of template matching on experimental subtomograms of ribosomes

With the same hypothesis testing procedure, we performed template matching on experimental subtomograms of Yeast 80S ribosome (PDB ID: 4V7R) from the EMPIAR-10045 database [18]. We used the 07

---

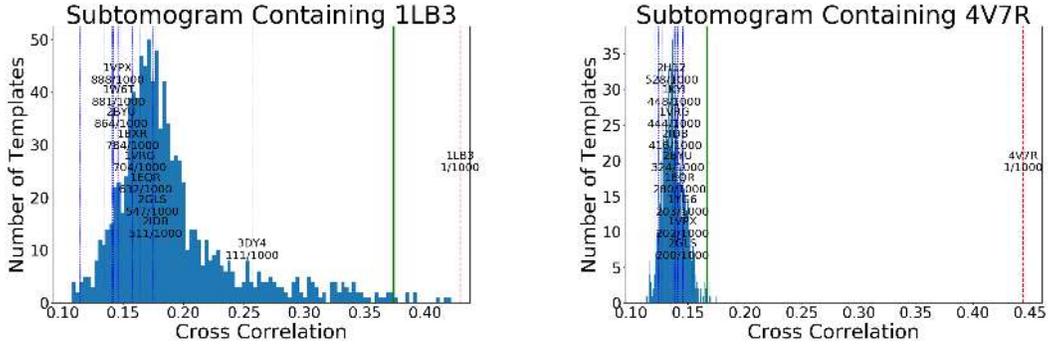

Figure 4: Successful hypothesis tests where $c(P_C, T(C_I))$ satisfies Cond. 1 and 2, and $C = C_I$. The blue histograms model $c(P_C, T(\tilde{C})), \tilde{C} \in \mathscr{S}_{pseudo}$. The vertical dashed lines denote $c(P_C, T(C')), C' \in \mathscr{S}_{known} \setminus \{C_I\}$, with the red line marking $c(P_C, T(C_I))$. The ten lines with the highest alignment scores are labeled with both their PDB IDs and their rank out of 1000 (i.e. the empirical p-value). The green line marks the alignment score at p-value threshold of 0.01.

tomogram, which contains 376 subtomograms in total. The experimental subtomograms were originally $200^3$ voxels with 0.217nm pixel size. For our experiments, they were preprocessed with a 2.17nm Gaussian blur and resized to $64^3$ voxels with 0.6nm pixel size, the same dimension and pixel size as the density maps of the training structures and the simulated subtomograms. As shown in Figure 5A, these subtomograms are visually much noisier than the simulated ones, template matching extremely challenging. Without preprocessing, it is difficult to even visually detect the macromolecular structure contained within the subtomograms.

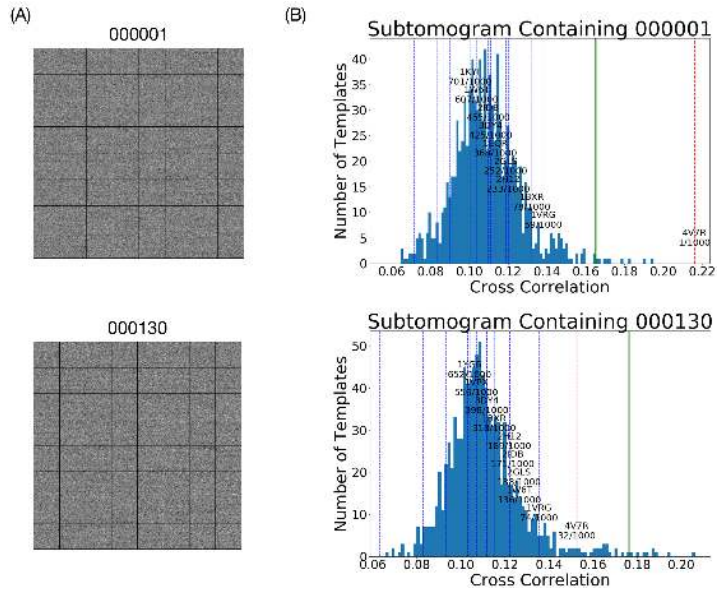

Figure 5: (A) Slices of raw subtomograms containing ribosome; (B) Hypothesis tests on experimental subtomograms of ribosome. We show both a successful and unsuccessful case.

With a p-value cut-off at 0.01, there were 264 (70.21%) successful tests, when Cond. 1 and Cond. 2 were met for the subtomogram $P$ and a ribosome complex of interest. 45 subtomograms (11.97%) resulted in Cond. 1 failure. 67 subtomograms (17.82%) resulted in Cond. 2 failure. Figure 5B shows one successful and a case of Cond. 2 failure from our experiments.

We note that due to our rescaling procedure, some subtomograms may be missing part of their ribosome structure, which may cause template matching to fail. Therefore, we further selected a subset of 100 subtomograms with the highest ribosome alignment score, corresponding to high confidence in containing the whole ribosome structure. Of these 100 subtomograms, 92 were successful, 1 resulted in Cond. 1 failure and 7 resulted in Cond. 2 failure.

### 3.4 Detecting False Positives

When the subtomogram $P$ does not contain a macromolecule or when the macromolecule in $P$ does not match with any of the known templates, it is desirable for template matching to conclude that no template matches with $P$. This is not possible if the template with the highest alignment scores is always concluded to be a match. Even when thresholding, it is difficult to choose a single cutoff that works in all possible cases and requires hyperparameter tuning. Using 20 experimental subtomograms that do not contain any macromolecules, we performed template matching along with our hypothesis testing procedure and we were able to prevent 40% of the false positives that would have occurred if we simply chose the highest alignment score as a match. Our statistical testing method makes reliable claims on this and drastically prevents the number of false positives.

### 3.5 Learning the shape manifold of macromolecules

Similar to previous works [33, 27] with GANs, we found that our model was able to capture the shape manifold of macromolecular complexes. We show this by interpolating between the latent vectors of $S$ and $D$, resulting in a smooth transition from a proteasome (PDB ID: 3DY4) to a ribosome (PDB ID: 4V7R) as shown in Figure 6. Our starting point $S$ was the latent vector of the nearest pseudo-complex for the proteasome, and our ending point $D$ was the latent vector of the nearest pseudo-complex for the ribosome. We generated the pseudo-complex at the $i$th step with the input vector $S + (D - S) \cdot i$. This smooth "deformation" from one structure to another illustrates the shape manifold of macromolecules and results in an effect similar to deformable image registration. Since deformable image registration can be quite computationally expensive (i.e. Large Deformation Diffeomorphic Metric Mapping [24]), with a well-trained model, our 3D-WGAN model could potentially be used as a computationally efficient heuristic for deformable registration of 3D shapes.

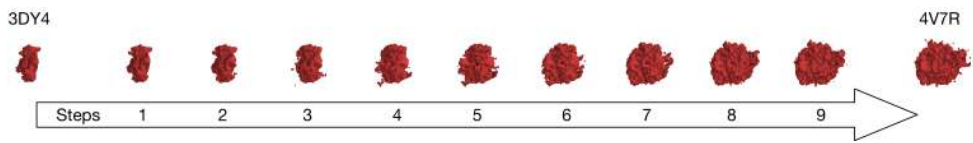

Figure 6: Intermediate shapes generated by interpolating between the latent representation of proteasome (PDB ID: 3DY4) and ribosome (PDB ID: 4V7R). This shows that our generative model can learn the shape manifold of macromolecular structures.

## 4 Conclusion

Without hypothesis testing, existing template matching approaches are not rigorous and are not statistically credible enough. Physical limitations to CECT remain a major difficulty that can bias the cross-correlation score and may even cause template matching to fail. To reliably conclude the identity of a subtomogram and to reduce false positive rates, we propose an image-derived approach for performing hypothesis testing for template matching by constructing a generative model for macromolecular complexes. We used the 3D-WGAN since it could efficiently produce a diverse population of novel pseudo-macromolecular complexes that are not simply rigid rotations of the original training structures. By sampling from the learned distribution of macromolecules $f_0$, we could successfully conclude with high statistical credibility that $P = T(C_I)$ with both simulated and experimental subtomograms. In addition, the 3D-WGAN generative model has potentially other applications in CECT. We have shown that the 3D-WGAN can learn the shape manifold of macromolecules [33]. By interpolating between latent representations, a 3D-WGAN can be used to visualize smooth transitions between structures. This ability for smooth deformations can be potentially extended to a heuristic for deformable image registration.

In addition, our approach is not limited to the 3D-WGAN and works with any reasonable generative model. A future work would include experimenting other generative approaches (e.g. VAE, shape-space modeling [24]) with the hypothesis testing procedure. The current GAN approach may also be improved with larger training sets or more complex architecture (e.g. [19]). Finally, our procedure of using generative modeling for hypothesis testing can be generalized to be applicable in many other template matching tasks with 2D or 3D images, such as object recognition and eye detection [29].

# 5 Acknowledgements

We thank Dr. Robert Murphy and Dr. Christopher Langmead for suggestions. This work was supported in part by U.S. National Institutes of Health (NIH) grant P41 GM103712. MX acknowledge support from Samuel and Emma Winters Foundation.

# 6 Supplementary Materials

## 6.1 Implementation details

A modified version of the Tomominer package [10] was used for processing the images. EMAN2 was used for back projection reconstruction. Mayavi was used for plotting the isosurfaces. Keras and Tensorflow were used for constructing and training the 3D-WGAN. The test is performed on a computer equipped with Intel i7 CPU, 128GB memory, and Nvidia GTX 1080 and GTX 1080 Ti GPUs.

## 6.2 Generating simulated subtomograms

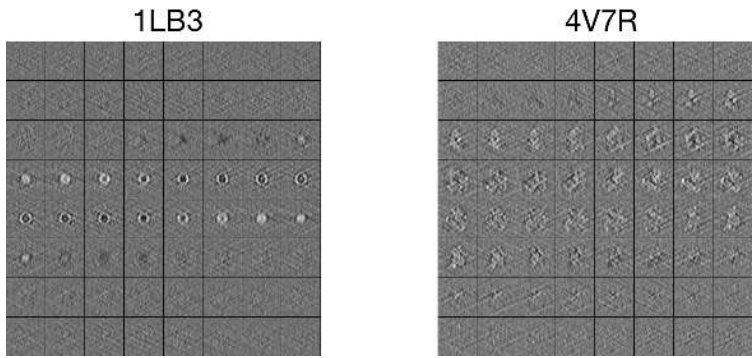

Figure 7: Raw slices of simulated subtomograms constructed from Mouse L-chain ferritin (PDB ID: 1LB3) and ribosome (PDB ID: 4V7R), with SNR of 0.03.

For our experiments on simulated subtomograms to be as reliable as possible, we simulated subtomograms by performing the actual tomographic image reconstruction process in a similar way as previous works [8, 4, 25, 35]. We properly included noise, and the missing wedge effect, and electron optical factors, such as the Contrast Transfer Function (CTF) and Modulation Transfer Function (MTF), assuming that the electron optical density of macromolecular complexes is proportional to the electrostatic potential. We used the PDB2VOL program from the Situs [32] package to generate volumes of $64^3$ voxels with a resolution and voxel spacing of 0.6nm. The density maps of known structure were used to simulate electron micrograph images using the tilt-angle of $\pm 60°$. We added noise to electron micrograph images [8] to achieve the desired SNR of 0.03, which was the lowest SNR for the fast alignment method to achieve successful alignment on all simulated subtomograms. Next, we convolved the electron micrograph images with the CTF and MTF to simulate optical effects [9, 25]. The acquisition parameters used are typical of those found in experimental tomograms [37], with spherical aberration of 2mm, defocus of -5$\mu$m, and voltage of 300kV. The MTF is defined as $\text{sinc}(\pi\omega/2)$ where $\omega$ is the fraction of the Nyquist frequency, which corresponds to a realistic detector [23]. Finally, a direct Fourier inversion reconstruction algorithm (implemented in the EMAN2 library [11]) is used to produce the simulated subtomogram according to the tilt-angle. Figure 7 shows examples of such simulated subtomograms.

The templates used for template matching were constructed using a tilt-angle range of ±90 degrees and infinite SNR, and with the same acquisition parameters used to simulate the subtomograms. The low SNR and missing wedge values make it very difficult to properly align subtomograms, especially smaller ones such as Ornithine carbamoyltransferase (PDB ID: 1A1S), but we found that including a Gaussian blur of 0.2nm increased the success rate of our fast alignment method.

### 6.3 Table of PDB ID

We collected 15 macromolecular complexes from the Protein Databank (PDB) [6], shown in the following table.

| PDB ID | Macromolecular Complex |
|--------|------------------------|
| 1A1S   | Ornithine carbamoyltransferase |
| 1BXR   | Carbamoyl phosphate synthetase |
| 1EQR   | Aspartyl tRNA-synthetase |
| 1KYI   | HslUV (H. influenzae)-NLVS Vinyl Sulfone Inhibitor Complex |
| 1LB3   | Mouse L-chain ferritin |
| 1VPX   | Transaldolase |
| 1VRG   | Propionyl-CoA carboxylase, beta subunit |
| 1W6T   | Octameric enolase |
| 1YG6   | ClpP |
| 2BYU   | Small heat shock protein Acr1 |
| 2GLS   | Glutamine synthetase |
| 2H12   | Acetobacter aceti citrate synthase |
| 2IDB   | 3-octaprenyl-4-hydroxybenzoate decarboxylase |
| 3DY4   | Yeast 20S proteasome |
| 4V7R   | Yeast 80S ribosome |

Table 1: The experimental macromolecular complexes used in this paper. They were used as training data for the 3D-WGAN, as complexes used to construct real templates and as complexes for simulation of subtomograms.

## 6.4 Details of the 3D-WGAN

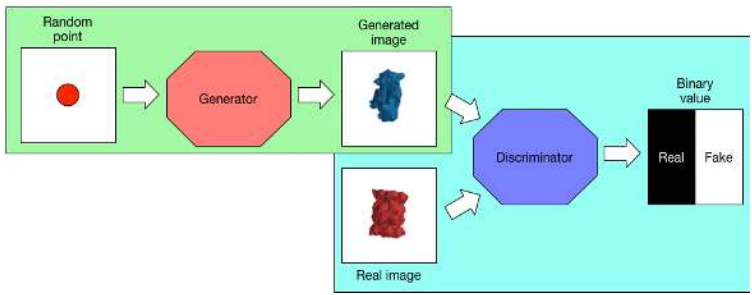

Figure 8: This diagram shows the concept of GAN.

Our generative approach uses techniques from deep learning, specifically the convolutional neural network and the GAN. Illustrated by Figure 8, the GAN involves a generator $G$ and a discriminator $D$. When given a sample of random points $z$. the goal of $G$ is to output its best attempt of sampling an image from the distribution on the manifold of training images. The discriminator model $D$ is trained to label generated (pseudo) images as 1 and training (real) images as 0. During training, we optimize $D$ to maximize its ability to classify real and pseudo images correctly, and we optimize $G$ to minimize the discrimination ability of $D$. This cost can be described by the binary cross-entropy equation [12, 33]

$$\min_G \max_D \ \mathbb{E}_{x \sim \mathbb{P}_{\text{data}}(x)} \log(D(x)) + \mathbb{E}_{z \sim \mathbb{P}_{\text{latent}}(z)} \log(1 - D(G(z))). \tag{5}$$

In our case, $x \sim \mathbb{P}_{\text{data}}(x)$ represents sampling from training data of real structures, and $z \sim \mathbb{P}_{\text{latent}}(z)$ represents sampling from the standard multidimensional Gaussian distribution in the latent space.

For our model, we used an improved and stabilized version of the GAN, called the Improved Wasserstein GAN (WGAN), which instead optimizes the Wasserstein distance, a cost function that is more favorable for optimization [1, 13]. Given two distributions $\mathbb{P}_1$ and $\mathbb{P}_2$, the Wasserstein distance can be calculated as

$$W(\mathbb{P}_r, \mathbb{P}_\theta) = \sup_{||f||_L \leq 1} \mathbb{E}_{x \sim \mathbb{P}_1}[f(x)] - \mathbb{E}_{x \sim \mathbb{P}_2}[f(x)], \tag{6}$$

according to [31], where the supremum is over all the 1-Lipschitz functions $f : \chi \to \mathbb{R}$. The Wasserstein distance describes the "distance" between two probability distributions. A benefit of using the Wasserstein distance is that it is continuous and almost differentiable everywhere. This allows us to train $D$ until convergence. As $D$ becomes more accurate, $G$ can learn to generate more realistic images. $G$ and $D$ continue to play the minimax game until they converge near to an optimum state, at which point the $G$ would produce diverse and realistic samples from the training image distribution.

The generator $G$ and the discriminator $D$ are implemented as convolutional neural networks (CNN). The CNN is a stack of convolution layers that can extract a complex hierarchy of image features, which has shown very successful results in computer vision applications such as object recognition and classification for both 2D and 3D images[20, 22]. [33] was the first attempt of using 3D CNNs with GANs and we based our network architecture around it. Although [33] uses five convolution layers, we found that four layers were enough to capture the structure of macromolecular complexes. In our 3D-WGAN generative model, we also use the transposed convolution layer [33] instead of the up-sampling layers to produce larger-size images. Each convolution layer is a collection of learned feature extractors [3]. Given N filters of size $K^3$ and stride S, the output y is the result of "sliding" each filter in steps of $S$, on top of the input and summing up the dot product at every location. Let the input to the layer be $X$ with $D$ volumetric slices, the filters of the layer be $W$ and the output be $y$. Mathematically [30], the convolutional layer is represented by

$$y_{i,j,k}^m = \sum_{p=0}^{D-1} \sum_{a=0}^{K-1} \sum_{b=0}^{K-1} \sum_{c=0}^{K-1} W_{a,b,c}^m \cdot X_{a+S \cdot i, b+S \cdot j, c+S \cdot k}^p \tag{7}$$

, where, $y_{i,j,k}^m$ is the index $(i,j,k)$ of the $m^{\text{th}}$ output volumetric slice , $W_{a,b,c}^m$ is the index $(a,b,c)$ of the $m^{\text{th}}$ filter, and $X_{a+S \cdot i, b+S \cdot j, c+S \cdot k}^p$ is the index $(a+S \cdot i, b+S \cdot j, c+S \cdot k)$ of the $p^{\text{th}}$ input volumetric slice. For example, a filter can learn to be like an edge extractor by learning weights similar to a Sobel kernel.

The Batch Normalization layer guarantees 0 mean and unit variance for inputs to hidden layers, which has been shown to stabilize training of deep networks [17]. Given a batch of inputs $x$, the batch normalized output is

$$\text{BN}(x) = \gamma \cdot \frac{x - \mathbb{E}[x]}{\sqrt{\text{Var}[x] + \varepsilon}} + \beta \tag{8}$$

, where $\gamma$ and $\beta$ are learned parameters, $\varepsilon$ is a small constant to prevent division by 0, and $\mathbb{E}[x]$ and $\text{Var}[x]$ are the mean and variance of $x$ respectively.

The activation functions used in our model are LeakyReLU and the hyperbolic tangent tanh. The LeakyReLU activation with a gradient of $m$ is described by Equation 9. The purpose of the Rectified Linear Unit (ReLU) and its variants (e.g. LeakyReLU) is to introduce non-linearity into the network, and have become widely used for its ability to speedup the convergence of deep networks [20, 15]. The hyperbolic tangent is described by Equation 10, and was used to bound the output of the generator within the range [-1, 1], as suggested by [27].

$$\text{LeakyReLU}(x) = \max(mx, x) \tag{9} \qquad \tanh(x) = \frac{e^x - e^{-x}}{e^x + e^{-x}} \tag{10}$$

A common problem in GANs that we also experienced was mode collapse. This collapse of the generator occurs when the generator produces very uniform and non-diverse structures. The minibatch discrimination layer was proposed by [28] to reduce the collapse of the generator by penalizing low-entropy generators. Intuitively, this layer allows the discriminator to observe many samples at once, so that it can also take entropy of a batch of samples into account when deciding between real and pseudo images. Following the method proposed in [28], introducing such a measure of "(low) entropy" is as follows. For a minibatch discrimination layer, we introduce a trainable tensor T, and let $f(X_i)$ be some features extracted from the sample $X_i$ (i.e. output of some intermediate layer). For every $f(X_i)$ s.t. $0 \le i < n$, compute the matrix $M_i = f(X_i)T$. Now, for every row $b$ of $M_i$, compare it with every $M_j$ by calculating

$$MB_b(X_i) = \sum_{j=0}^{n} \exp(-|M_{i,b} - M_{j,b}|) \tag{11}$$

, and $MB(X_i) = [MB_1(X_i), MB_2(X_i), ..., MB_b(X_i)]$. These outputs define a measure of "(low) entropy". We concatenate $MB(X) = [MB(X_1), MB(X_2), ..., MB(X_n)]$ to the output of some intermediate layer, so that the discriminator can take this measure of entropy into account.

## 6.5 Cases of failed simulated hypothesis tests

In our simulated tests, on average, 6.67% of our tests resulted in Cond. 1 failure and 10.67% of our tests resulted in Cond. 2 failure.

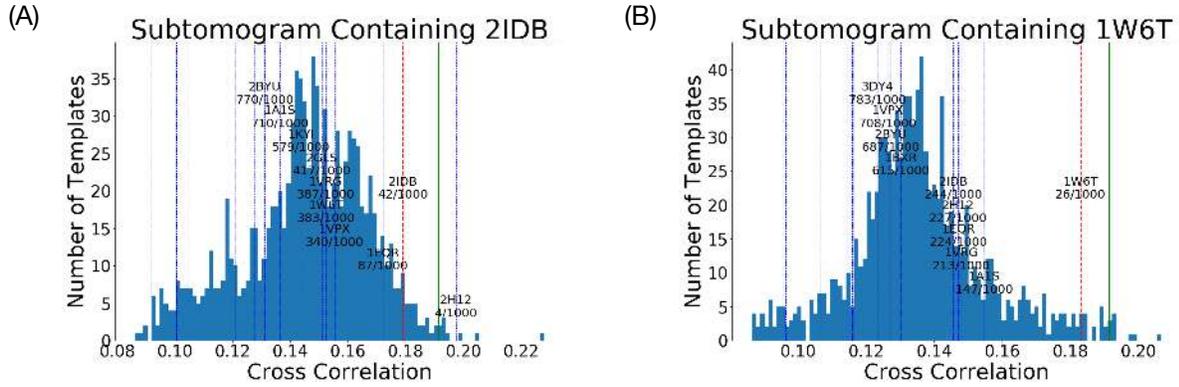

Figure 9: Examples of unsuccessful results. (A) is a Cond. 1 failure likely caused by bias in alignment due to low SNR; (B) is a Cond. 2 failure as it has a p-value above 0.01.

## 6.6 Supplementary Figures

### 6.6.1 Simulated Results

The following two pages are the remainder of the hypothesis tests on simulated subtomograms. All of these are successful with the exception of 1VPX.

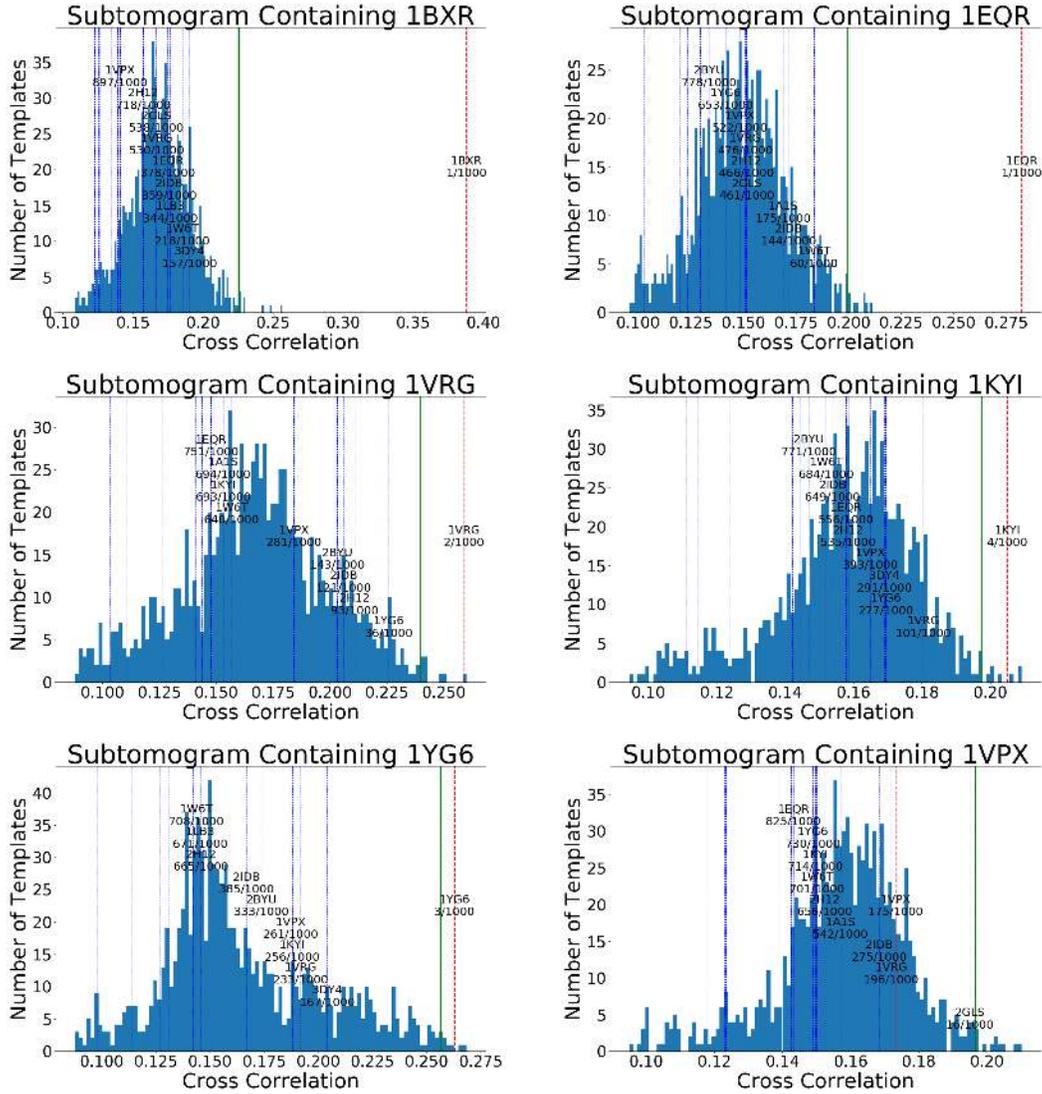

Figure 10: First batch of the remainder of the hypothesis tests on simulated subtomograms. Subtomograms were simulated using default parameters discussed in Appendix 6.2; 0.03 SNR with 0.2nm smoothing.

Figure 11: Second batch of the remainder of the hypothesis tests on simulated subtomograms. Subtomograms were simulated using default parameters discussed in Appendix 6.2; 0.03 SNR with 0.2nm smoothing.